\documentclass[apj]{emulateapj}

\usepackage{multirow}
\usepackage{color} 

\usepackage{lineno}

\def\ergs{erg~s$^{-1}$}
\def\ergcms{erg~cm$^{-2}$~s$^{-1}$}

\def\NuSTAR{\textit{NuSTAR}}
\def\Swift{\textit{Swift}}
\def\RXTE{\textit{RXTE}}
\def\Chandra{\textit{Chandra}}
\def\BeppoSAX{\textit{BeppoSAX}}

\begin{document}
\title{\NuSTAR\ and $\Swift$ observations of the black hole candidate XTE J1908+094 during its 2013 outburst}

\author{Lian Tao\altaffilmark{1,2,3}, John A. Tomsick\altaffilmark{4}, Dominic J. Walton\altaffilmark{5,1}, Felix F{\"u}rst\altaffilmark{1}, Jamie Kennea\altaffilmark{6}, Jon M. Miller\altaffilmark{7},  Steven E. Boggs\altaffilmark{4}, Finn E. Christensen\altaffilmark{8}, William W. Craig\altaffilmark{4},
Poshak Gandhi\altaffilmark{9,10}, Brian W. Grefenstette\altaffilmark{1}, Charles J. Hailey\altaffilmark{11}, Fiona A. Harrison\altaffilmark{1}, Hans A. Krimm\altaffilmark{12,13}, Katja Pottschmidt\altaffilmark{12,14}, Daniel Stern\altaffilmark{5}, Shriharsh P. Tendulkar\altaffilmark{1}, and William W. Zhang\altaffilmark{12}}

\altaffiltext{1}{Cahill Center for Astronomy and Astrophysics, California Institute of Technology, Pasadena, CA 91125, USA}
\altaffiltext{2}{Department of Physics, Tsinghua University, Beijing 100084, China}
\altaffiltext{3}{Center for Astrophysics, Tsinghua University, Beijing 100084, China}
\altaffiltext{4}{Space Sciences Laboratory, University of California, Berkeley, CA 94720, USA}
\altaffiltext{5}{Jet Propulsion Laboratory, California Institute of Technology, Pasadena, CA 91109, USA}
\altaffiltext{6}{Department of Astronomy \& Astrophysics, The Pennsylvania State University, University Park, PA 16802, USA}
\altaffiltext{7}{Department of Astronomy, University of Michigan, 500 Church Street, Ann Arbor, MI 48109-1042, USA}
\altaffiltext{8}{DTU Space, National Space Institute, Technical University of Denmark, Elektrovej 327, DK-2800 Lyngby, Denmark}
\altaffiltext{9}{School of Physics \& Astronomy, University of Southampton, Highfield, Southampton SO17 1BJ, UK}
\altaffiltext{10}{Department of Physics, Durham University, South Road, Durham DH1 3LE, UK}
\altaffiltext{11}{Columbia Astrophysics Laboratory, Columbia University, New York, NY 10027, USA}
\altaffiltext{12}{NASA Goddard Space Flight Center, Greenbelt, MD 20771, USA}
\altaffiltext{13}{USRA, 10211 Wincopin Circle, Suite 500, Columbia, MD 21044, USA}
\altaffiltext{14}{CRESST, Department of Physics, and Center for Space Science and Technology, UMBC, Baltimore, MD 21250, USA}

\shorttitle{XTE J1908+094}
\shortauthors{Tao et al.}

\begin{abstract}
The black hole candidate XTE J1908+094 went into outburst for the first time since 2003 in October 2013. We report on an observation with the \textit{Nuclear Spectroscopic Telescope Array} ($\NuSTAR$) and monitoring observations with $\Swift$ during the outburst. $\NuSTAR$ caught the source in the soft state: the spectra show a broad relativistic iron line, and the light curves reveal a $\sim$40\,ks flare with the count rate peaking about 40\% above the non-flare level and with significant spectral variation. A model combining a multi-temperature thermal component, a power-law, and a reflection component with an iron line provides a good description of the $\NuSTAR$ spectrum.  Although relativistic broadening of the iron line is observed, it is not possible to constrain the black hole spin with these data.  The variability of the power-law component, which can also be modeled as a Comptonization component, is responsible for the flux and spectral change during the flare, suggesting that changes in the corona (or possibly continued jet activity) are the likely cause of the flare.

\end{abstract}


\section{Introduction}

XTE J1908+094 is an X-ray transient serendipitously discovered with the \textit{Rossi X-ray Timing Explorer} ($\RXTE$) Proportional Counter Array (PCA) when it went into outburst in 2002 February \citep{woo02}. The source flux in the 2--10\,keV band rose by a factor of $\sim$3 in one month \citep{woo02} and reached about 100 mCrab on 2002 April 6 \citep{gog04}. The power density spectrum showed a broad quasi-periodic oscillation (QPO) at 1\,Hz without any coherent pulsation between 0.001 and 1024\,Hz \citep{woo02}. In the X-ray energy spectrum, an iron emission line and a hard tail up to 250\,keV were detected \citep{woo02, fer02}. The hard X-ray spectrum did not agree with an extrapolation of the absorbed power-law model in the 2--30\,keV band reported by \citet{woo02}, and showed a high-energy cut-off at $\sim$ 100\,keV \citep{fer02}. Given the timing and spectral characteristics, XTE J1908+094 is suggested to be a black hole candidate \citep{woo02, fer02, int02, gog04}. 

The broad band X-ray spectrum of XTE J1908+094 is well fitted with two continuum components, a multi-temperature disk blackbody with $kT$ $\sim$ 0.8\,keV and a Compton plasma with a temperature near 40\,keV, and an emission line centered on the location of the Fe K$\alpha$ line \citep{int02}. The emission feature is very broad with FWHM$=3.2\pm0.5$\,keV (line width $\sigma=1.4\pm0.2$\,keV), which may be due to Compton scattering in a corona or the relativistic effects from gravitational redshift and Doppler broadening of orbital motion \citep{int02}. In the latter case, the broadening of the Fe emission line would be expected to be asymmetric, and could be used to measure the black hole (BH) spin \citep{rey03,mil07}. Based on this method, \citet{mil09} measured the dimensionless spin of XTE J1908+094 to be $a=0.75\pm 0.09$.

The radio counterpart of XTE J1908+094 was discovered with the Very Large Array (VLA) at R.A.=$19^{\rm h}08^{\rm m}53.\!^{\rm s}077$, Decl.=+$09^{\circ}23^{\prime}04.\!^{\prime\prime}90$ (J2000.0) \citep{rup02}, which is consistent with the $\Chandra$ position, R.A.=$19^{\rm h}08^{\rm m}53.\!^{\rm s}07$, Decl.=+$09^{\circ}23^{\prime}05.\!^{\prime\prime}0$ \citep{jon04}. \cite{cha02} identified a likely near-infrared (NIR) counterpart for the source, but the possible counterpart was resolved into two sources separated by $\sim$ 0.8\,arcsec in subsequent observations \citep{cha06}.  Both potential counterparts are consistent with XTE J1908+094 being a low mass X-ray binary (LMXB). One of the NIR sources would indicate an intermediate/late type (A-K) main-sequence companion star, while the other would indicate a late-type main-sequence companion star with spectral type later than K \citep{cha06}. Recently, $\Swift$/UVOT observed the X-ray source; however, no counterpart was found in the $V$ band with the 3$\sigma$ limiting magnitude to be $V > 20.3$ \citep{kri13b}. 

The distance to XTE J1908+094 is not well established. Based on its X-ray flux, the source is suggested to be at a distance greater than 3\,kpc \citep{int02}.  From the optical measurements, the possible distance range is 3--10\,kpc \citep{cha06}. An estimate using the X-ray and radio fluxes puts the source at a distance of $\sim$ 2--10\,kpc \citep{mil13}.

Previously, XTE J1908+094 went through two outbursts in 2002 and early 2003 with very similar spectral evolution \citep{gog04}. On 2013 October 26, another outburst of XTE J1908+094 was detected by the $\Swift$ Burst Alert Telescope (BAT), with the 15--50\,keV flux reaching $\sim$ 60\,mCrab two days later \citep{kri13a}. Subsequently, a number of telescopes, including $\NuSTAR$, $\Swift$ \citep{kri13a, kri13b}, the VLA \citep{mil13}, the Arcminute Microkelvin Imager (AMI) Large Array \citep{rus13a}, the \textit{Monitor of All-sky X-ray Image} \citep[\textit{MAXI},][]{neg13} and the Australia Telescope Compact Array \citep[ATCA,][]{cor13} carried out observations of the source. In this paper, we report on the $\NuSTAR$ and $\Swift$ observations of the 2013 outburst in detail (Section~\ref{sec:obs}) and investigate its spectral evolution and properties (Section~\ref{sec:res}). We present a discussion of the results in Section~\ref{sec:dis} and conclusions in Section~\ref{sec:sum}.   

\section{Observations}
\label{sec:obs}

XTE J1908+094 was monitored with many short observations by the $\Swift$/X-ray Telescope (XRT) \citep{bur05} from 2013 October 26 to 2013 December 3, and $\NuSTAR$ \citep{har13} carried out an observation with an effective exposure time of $\sim$ 45\,ks on 2013 November 8.  In order to study the outburst properties of the source, we used the $\NuSTAR$ observation and all of the $\Swift$/XRT observations that were long enough to achieve sufficient statistical quality (see Table~\ref{tab:obs} for the observation list).

\begin{deluxetable}{ccl}
\tablecolumns{3}
\tablewidth{0pc}
\tabletypesize{\scriptsize}
\tablecaption{$\NuSTAR$ and $\Swift$ Observations \label{tab:obs}}
\tablehead{
\colhead{ObsID}  & \colhead{Observed date} & \colhead{Exposure (s)} 
}
\startdata
\noalign{\smallskip}
\multicolumn{3}{l}{$\NuSTAR$}\\
\noalign{\smallskip}
80001014002 & 2013-11-08 & 45061/45404$^{a}$\\
\noalign{\smallskip}
\multicolumn{3}{l}{$\Swift$ XRT}\\
\noalign{\smallskip}
00033014001 & 2013-10-29  & 972  \\
00033014002 & 2013-11-01  & 1224 \\
00033014003 & 2013-11-03  & 1032 \\
00033014004 & 2013-11-08  & 1044 \\
00033014005 & 2013-11-09  & 1007 \\
00033014006 & 2013-11-10  & 503 \\
00033014007 & 2013-11-11  & 880 \\
00033014008 & 2013-11-12  & 975 \\
00033014009 & 2013-11-13  & 1343 \\
00033014010 & 2013-11-14  & 792 \\
00033014011 & 2013-11-15  & 1085 \\
00033014012 & 2013-11-16  & 1150 \\
00033014014 & 2013-11-18  & 979 \\
00033014015 & 2013-11-19  & 1008 \\
00033014016 & 2013-11-20  & 978 \\
00033014017 & 2013-11-23  & 958 \\
00033014018 & 2013-11-28  & 824 \\
00033014019 & 2013-12-03  & 976
\enddata
\tablecomments{$^{a}$ The exposure times of $\NuSTAR$ FPMA and FPMB, respectively. The $\Swift$ XRT data are taken in windowed mode.}
\end{deluxetable}

\subsection{$\NuSTAR$}
The $\NuSTAR$ data (ObsID 80001014002) were processed using version 1.3.1 of the NuSTARDAS pipeline with $\NuSTAR$ CALDB version 20131223. The spectra and light curves were extracted from a region centered at the position of XTE J1908+094 with a radius of 120$\arcsec$. The source region was contaminated by stray light from the nearby bright source GRS 1915+105.  Thus, the background region was chosen carefully. We used a circular background region with a radius of 80$\arcsec$ from the part of the field of view that was illuminated by the GRS 1915+105 stray light and as far away from XTE J1908+094 as possible. The background count rate is less than 6\% of the source count rate, which means that, even considering the stray light, the source still strongly dominates the spectra and light curves.  The spectra of the two $NuSTAR$ focal plane modules A and B (FPMA and FPMB), were rebinned to have at least 50 counts per bin. The light curves were binned to a time resolution of 100\,s.

\subsection{$\Swift$}

We reduced the $\Swift$/XRT data from 2013 October 29 to 2013 December 3 (see Table~\ref{tab:obs}). All data were taken in windowed timing mode. Using XSELECT with XRT CALDB version 20140709, the spectra were extracted from a circular region with a radius of 20 pixels ($\sim 47\arcsec$). The background extraction region is a box 20 pixels long, centered 100 pixels from the middle of the source extraction region. Ancillary response files were created using the ftool \texttt{xrtmkarf}. At lower energies, the windowed timing mode shows a bump between 0.4-1 keV and a turn up at the lowest energies\footnote{http://www.swift.ac.uk/analysis/xrt/digest\_cal.php}. In order to reduce the low-energy spectral residuals, the grade 0 data and the position-dependent response matrices\footnote{http://www.swift.ac.uk/analysis/xrt/rmfs.php} from the latest XRT calibration files were used. Finally, the extracted spectra were rebinned to contain a minimum of 25 counts per bin.

\begin{figure}
\centering
\includegraphics[width=0.45\textwidth]{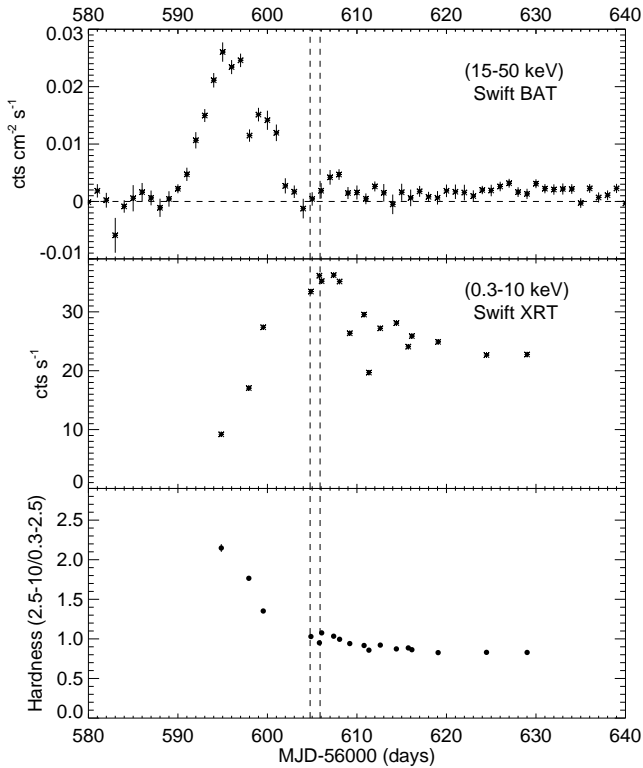}
\caption{From top to bottom, the 15-50 keV $\Swift$/BAT count rate, the 0.3-10 keV $\Swift$/XRT count rate and the hardness ratio between the $\Swift$/XRT hard band (2.5-10 keV) and soft band (0.3-2.5 keV). Two vertical dashed lines indicate the time boundaries of the $\NuSTAR$ observation.
\label{fig:cts}}
\end{figure}

\begin{figure}
\centering
\includegraphics[width=0.48\textwidth]{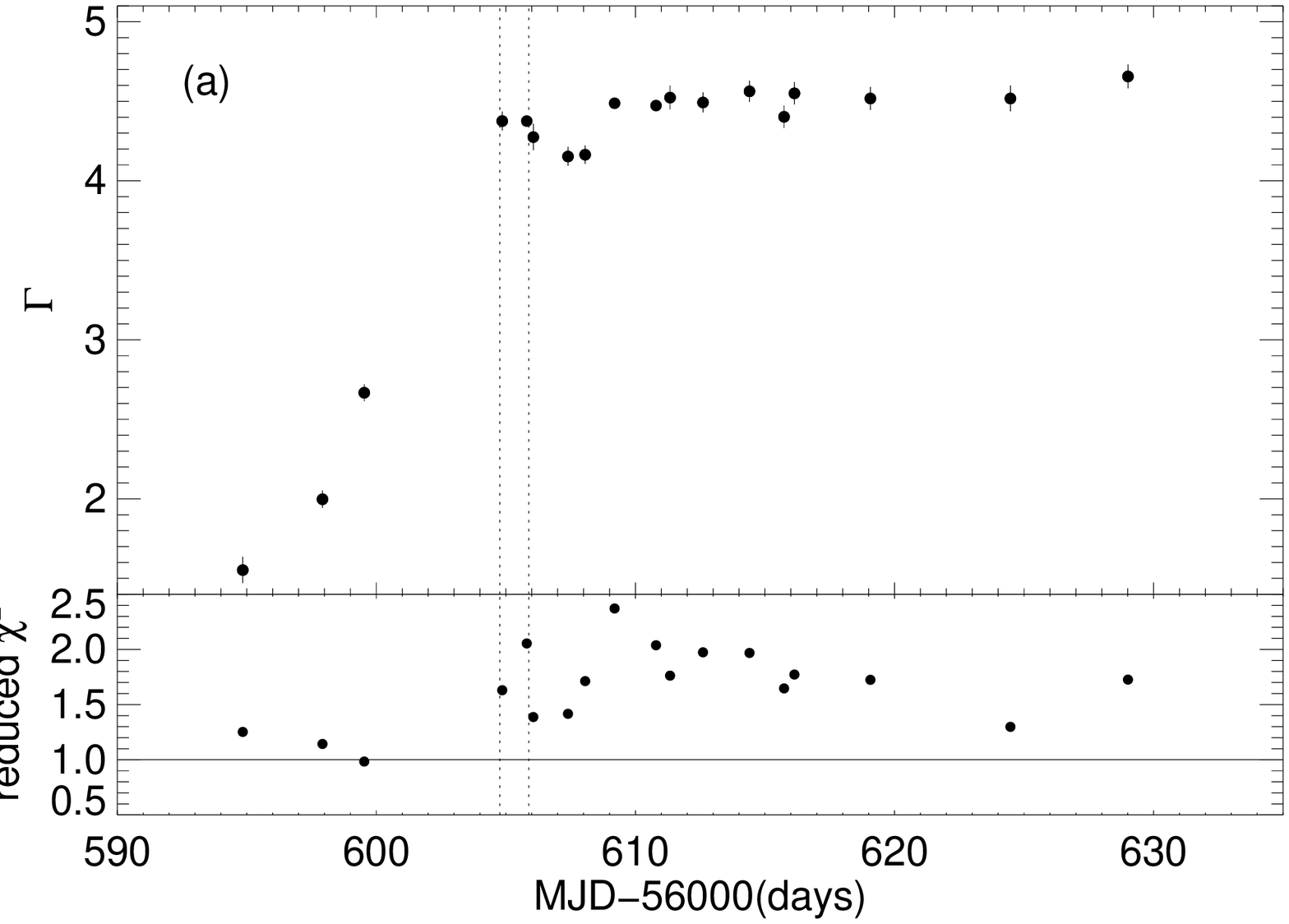}\\
\includegraphics[width=0.48\textwidth]{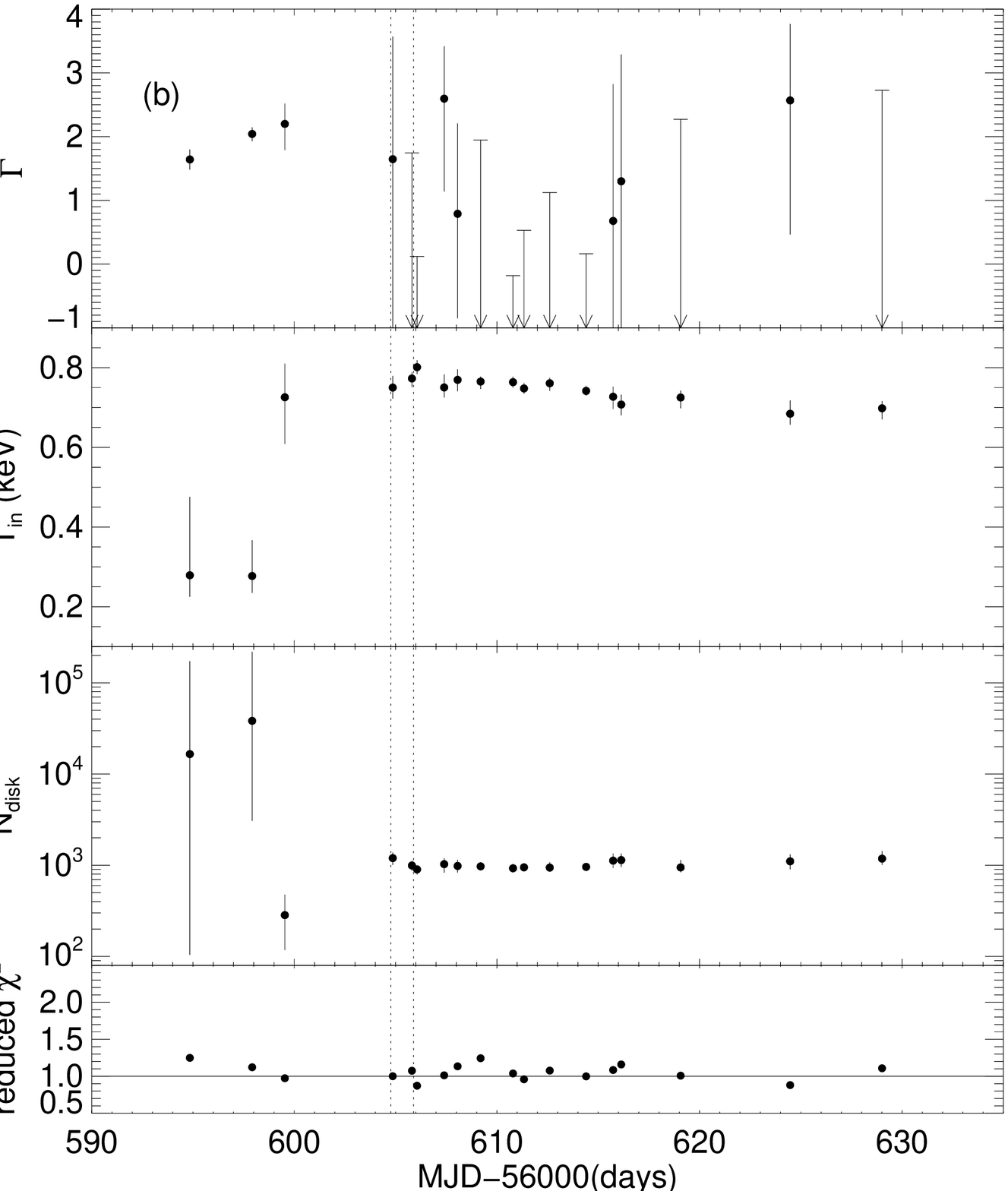}
\caption{The evolution of $\Swift$/XRT spectral parameters. (a) photon index $\Gamma$ and reduced $\chi^2$ when fitting with a single absorbed power-law model; (b) photon index $\Gamma$, inner disk temperature $T_{\rm in}$, normalization of the {\tt diskbb} model and reduced $\chi^2$ when fitting with a two-component model consisting of power-law and disk components. The arrows indicate the upper limit of $\Gamma$ where the lower error bars of $\Gamma$ could not be well constrained. Two vertical dotted lines show the time interval of the $\NuSTAR$ observation.
\label{fig:para}}
\end{figure}

\begin{figure}
\centering
\includegraphics[width=0.45\textwidth]{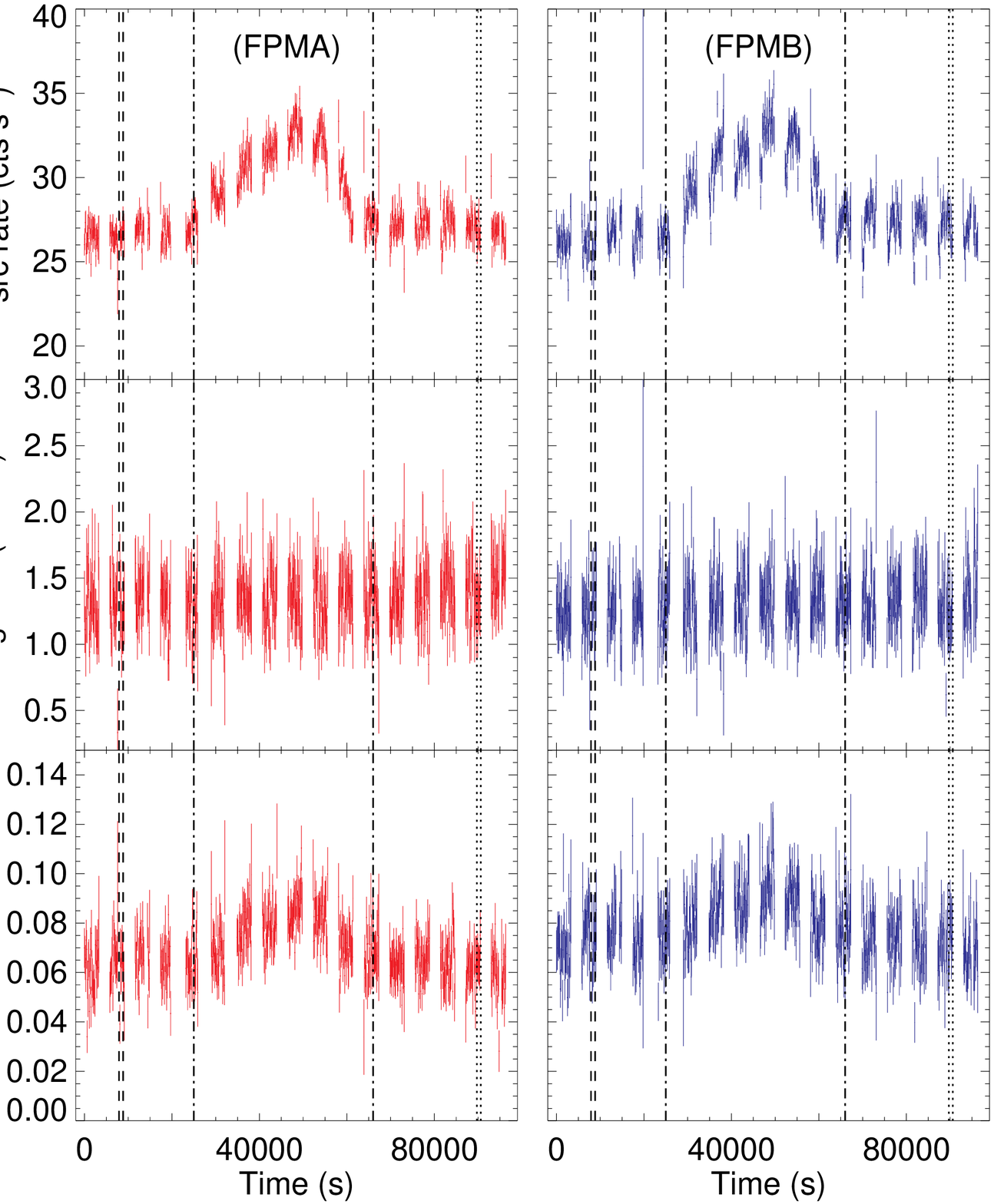}
\caption{The $\NuSTAR$ observation of XTE J1908+094. Top: the 3--79 keV light curves of XTE J1908+094 observed by FPMA and FPMB, respectively. Middle: the background light curves in the 3--79 keV band. Bottom: the hardness ratio defined as the ratio of the count rates in the 10--79 keV to 3--10 keV bands. The two vertical dash-dotted lines exhibit the duration of the flare, the two vertical dashed lines show the time interval of $\Swift$ observation ObsID 00033014004, and the two vertical dotted lines indicate $\Swift$ observation ObsID 00033014005. 
\label{fig:bum}}
\end{figure}

\section{Results}
\label{sec:res}

\subsection{$\Swift$ Monitoring}
\label{sec:swi}

The $\Swift$ monitoring observations reveal a clear evolution starting from 2013 October 25 (MJD 56590) (see Figure~\ref{fig:cts}). The $\Swift$/BAT count rate in the 15--50\,keV band\footnote{Available: http://swift.gsfc.nasa.gov/results/transients/weak/XTEJ1908p094} increased rapidly from $0.0022 \pm 0.0008$ $\rm cts~cm^{-2}~s^{-1}$ on MJD 56590 to $0.026 \pm 0.002$ $\rm cts~cm^{-2}~s^{-1}$ on MJD 56595 and then decreased sharply to $\sim 0.0015$ $\rm cts~cm^{-2}~s^{-1}$ and stayed close to that level after MJD 56604. In $\Swift$/XRT's 0.3--10\,keV band, the source brightened from $9.2 \pm 0.1$ $\rm cts~s^{-1}$ on MJD 56595, reaching its peak count rate of $36.2 \pm 0.2$ $\rm cts~s^{-1}$ on MJD 56607 and then dimmed.  The hardness, defined as the ratio of the count rates in the 2.5--10\,keV to 0.3--2.5\,keV bands, started to decrease from $2.15 \pm 0.05$ on MJD 56595 to $1.030 \pm 0.011$ on MJD 56605, and then stayed at a value of $\sim 1$. All of these measurements suggest that the source entered a state transition around MJD 56595 and was in the soft state 10 days later. The long exposure obtained with $\NuSTAR$ between MJD 56605 and MJD 56606 occurred after the source reached the soft state.

First, we fitted the 0.5--10\,keV $\Swift$/XRT spectra using a single absorbed power-law model. The $\Swift$ data below 0.5\,keV were ignored during the spectral fits in order to exclude the low-energy spectral residuals in windowed timing mode. The values of the photon index, $\rm \Gamma$, and the reduced $\chi^2$ are plotted in Figure~\ref{fig:para}a. The value of $\Gamma$ increased steeply from 1.6 on MJD 56595 to 4.4 on MJD 56605, and remained at $\sim 4.5$ until MJD 56629, consistent with the source going through the hard to soft state transition.  After the source begins the state transition, the accretion disk is significant for most of the observation.  For these observations, a single power-law does not provide a good fit to the spectra, and the addition of a disk-blackbody component provides a significant improvement to the fit.  The inner disk temperature $T_{\rm in}$, the normalization of the {\tt diskbb} model, the photon index $\Gamma$, and the reduced $\chi^2$ are shown in Figure~\ref{fig:para}b. The absorbed disk blackbody plus power-law model could successfully fit all spectra, with $T_{\rm in}$ increasing from 0.3 before the state transition and stabilizing at about 0.7--0.8\,keV in soft state.

\begin{figure}
\centering
\includegraphics[width=0.48\textwidth]{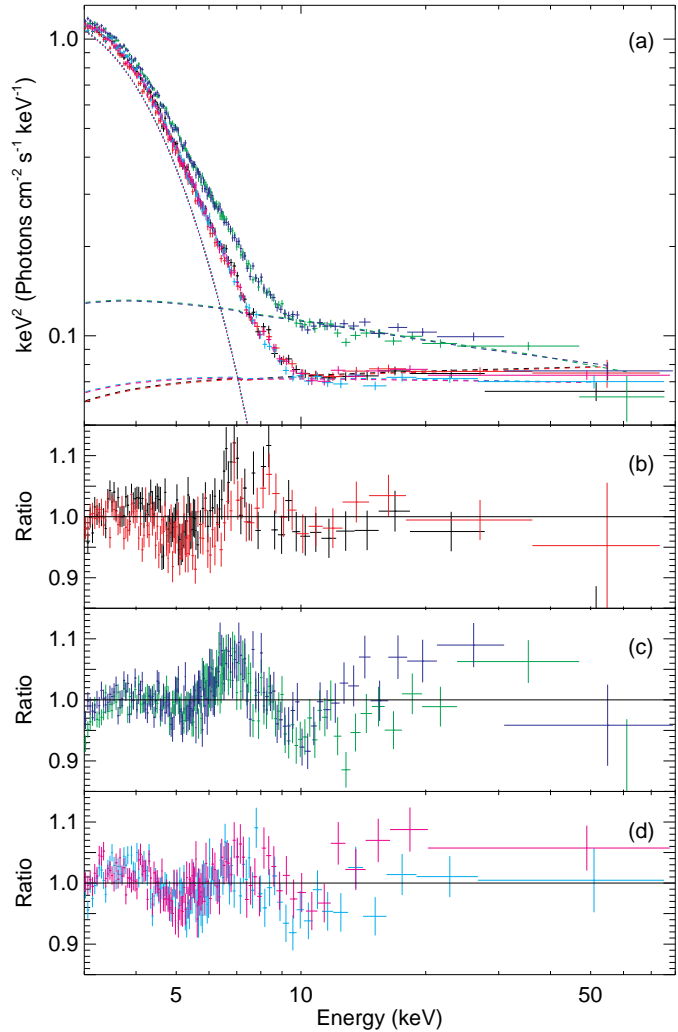} 
\caption{The $\NuSTAR$ spectra, model and residuals. (a) the unfolded $\NuSTAR$ spectra and model in $E^2\ast f(E)$ representation. The model includes a fixed disk blackbody component and a free power-law model before, during and after the flare. Black and red crosses are $\NuSTAR$ FPMA and FPMB spectra before the flare, respectively; green and blue crosses indicate the spectra during the flare; cyan and magenta crosses indicate the spectra after the flare. The top dotted line exhibits the disk blackbody component, and the three lower dashed lines show the power-law components during the different stages. The power-law component is stronger during the flare, while the power-law components before and after the flare show fluxes that are lower and similar to each other. Also, the power-law index of the flare is softer than those of the non-flare spectra. (b) -- (d) Data/model ratio before, during and after the flare, respectively.
\label{fig:mo1}}
\end{figure}

\subsection{$\NuSTAR$ Spectroscopy}

The $\NuSTAR$ light curves of FPMA and FPMB (see the top panels of Figure~\ref{fig:bum}) with background subtraction show a flare of $\sim 40$ ks duration with the peak rate being $\sim$40\% above the non-flare rate. The background light curves are also shown in Figure~\ref{fig:bum} in order to evaluate if the variability might be from the nearby source GRS 1915+105 rather than XTE J1908+094. The background light curves are stable at an average value of 1.3 $\rm cts~s^{-1}$, less than 6\% of the net source count rate. Thus, although the high background caused by GRS 1915+105 affects the statistical quality of the XTE J1908+094 spectrum, Figure~\ref{fig:bum} demonstrates that the flare in the light curves comes from XTE J1908+094. To study whether the flare has a different spectrum from the non-flare emission, we first checked the ratios of the 10--79 keV count rates to the 3-10 keV count rates. During the flare, this hardness ratio increased (bottom panels of Figure~\ref{fig:bum}), indicating that there is spectral variation.  

To investigate further, we extracted the 3--79\,keV spectra prior to the flare, during the flare and after the flare and fitted them together using a simple model combining an energy-independent multiplicative factor ({\tt constant}), an absorption model ({\tt tbabs}), adopting abundances from \citet{wil00}, a power-law model ({\tt pegpwrlw}) and a multi-temperature disk-blackbody model ({\tt diskbb}), i.e, {\tt constant} $\ast$ {\tt tbabs} $\ast$ ({\tt pegpwrlw} + {\tt diskbb}). Untying the model parameters individually or in combination, we found that only changing the power-law model could explain the variability, with a reduced $\rm \chi^2=1.25$ for 2697 degrees of freedom (dof). As shown in the top panel of Figure~\ref{fig:mo1}, the power-law component changes significantly between the flare and non-flare spectra: before the flare, $\Gamma =1.96$, during the flare, $\Gamma =2.23$, and after the flare, $\Gamma =2.03$. Here, we quote the best fit parameters without error bars because this simple model does not provide an acceptable fit to the data. Moreover, as shown in Figure~\ref{fig:mo1}b, c and d, all spectra exhibit similar residuals when the power-law parameters are allowed to vary prior to the flare, during the flare and after the flare. Very poor fits are obtained if the power-law component is required to be the same for all three spectra. All of this suggests that the corona, rather than the mass accretion rate and the accretion disk, went through great changes during the $\NuSTAR$ observation.

\begin{figure}
\centering
\includegraphics[width=0.40\textwidth]{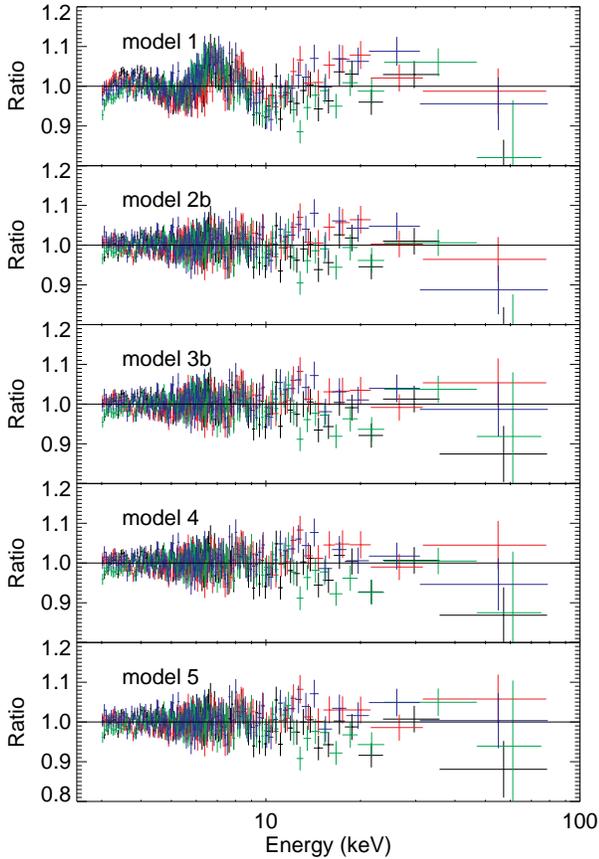} \\
\caption{The residuals for the best fit in different models. Black and red symbols are $\NuSTAR$ FPMA and FPMB spectra in the non-flare state; green and blue symbols are the spectra of the flare.   
\label{fig:mo3}}
\end{figure}

\begin{deluxetable*}{clllcllccccl}
\tablewidth{\textwidth}
\tablecolumns{12}
\setlength{\tabcolsep}{0pt}
\tabletypesize{\scriptsize}
\tablecaption{Spectral fitting of XTE J1908+094: part I  \label{tab:mo1}}
\tablehead{
\colhead{\multirow{2}*{Model}} & \colhead{\multirow{2}*{$C$}} & \colhead{\multirow{2}*{$N_{\rm H}$}} &\colhead{$\Gamma_1$} &\colhead{$N_{\rm PL1}$} & \colhead{$E_{\rm cent1}$} &\colhead{$\sigma_1$} & \colhead{$N_{\rm gauss1}$} & \colhead{\multirow{2}*{$kT$}} & \colhead{\multirow{2}*{$N_{\rm disk}$}} & \colhead{\multirow{2}*{$\chi^2/{\rm dof}$}} \\
\colhead{ } & \colhead{ } & \colhead{ } &\colhead{$\Gamma_2$} &\colhead{$N_{\rm PL2}$} & \colhead{$E_{\rm cent2}$} &\colhead{$\sigma_2$} & \colhead{$N_{\rm gauss2}$} & \colhead{ } & \colhead{ } & \colhead{ }
}
\startdata
\multirow{2}*{1} & \multirow{2}*{$0.993$} & \multirow{2}*{$1.6$} & $2.00$ & $385$ & \nodata &  \nodata & \nodata & \multirow{2}*{$0.783$} & \multirow{2}*{$650$} &  \multirow{2}*{2709.8/2099}  \\
  &  &   &  $2.23$ &  $554$ & \nodata & \nodata & \nodata &  &  &  \\
\noalign{\smallskip}\hline\noalign{\smallskip}
\multirow{2}*{2a} & \multirow{2}*{$0.993\pm0.003$} & \multirow{2}*{$2.5^{\star}$} & $1.95\pm0.03$ & $387\pm4$ & $6.82^{+0.14}_{-0.16}$ & $0.98^{+0.15}_{-0.14}$ & $0.74^{+0.17}_{-0.14}$ & \multirow{2}*{$0.755\pm0.003$} & \multirow{2}*{$873^{+20}_{-19}$} &   \multirow{2}*{2321.7/2094}  \\
  &  &   &  $2.15\pm0.03$ &  $545^{+5}_{-6}$ & $6.1^{+0.3}_{-0.4}$ & $1.3^{+0.3}_{-0.2}$ & $2.2^{+1.0}_{-0.6}$ &  &  &  \\
\noalign{\smallskip}\hline\noalign{\smallskip}
\multirow{2}*{2b} & \multirow{2}*{$0.993\pm0.003$} & \multirow{2}*{$5.8^{+0.8}_{-0.6}$} & $1.99\pm0.03$ & $391\pm4$ & $5.3^{+0.6}_{-1.0}$ &  $1.6^{+0.3}_{-0.2}$ & $5^{+7}_{-2}$ & \multirow{2}*{$0.671^{+0.018}_{-0.026}$} & \multirow{2}*{$2300^{+800}_{-400}$} &   \multirow{2}*{2207.8/2093}  \\
  &  &   &  $2.16\pm0.03$ &  $551\pm6$ & $4.9^{+0.5}_{-0.9}$ & $1.8^{+0.3}_{-0.2}$ & $10^{+8}_{-4}$ &  &  & 
\enddata
\tablecomments{\\
Model 1: {\tt constant} $\ast$ {\tt tbabs} $\ast$ ({\tt pegpwrlw} + {\tt diskbb}). \\
Model 2: {\tt constant} $\ast$ {\tt tbabs} $\ast$ ({\tt pegpwrlw} + {\tt gaussian} + {\tt diskbb}); 2a: fixed $N_{\rm H}$ at $2.5\times 10^{22}~\rm cm^{-2}$; 2b: $N_{\rm H}$ was set as a free parameter.\\
Model 1 is not an acceptable fit to the spectrum so we just quote the best fit parameters without error bars. \\
$^{\star}$: fixed value; \\
$C$ is the $\NuSTAR$ FPMB normalization factor relative to FPMA; \\
$N_{\rm H}$ is the X-ray absorption column density in units of $10^{22}~\rm cm^{-2}$; \\
$\Gamma_1$ and $\Gamma_2$ are the power-law photon indices of the non-flare and flare spectra; \\
$N_{\rm PL1}$ and $N_{\rm PL2}$ are the power-law component flux normalizations over the 3--79 keV energy band in units of $\rm 10^{-12}$ \ergcms; \\
$E_{\rm cent1}$ and $E_{\rm cent2}$ are the Gaussian emission line energies in keV;\\ 
$\sigma_1$ and $\sigma_2$ are the line widths in keV; \\
$N_{\rm gauss1}$ and $N_{\rm gauss2}$ are the Gaussian component normalizations in units of $\rm10^{-3}~photons~cm^{-2}~s^{-1}$; \\
$kT$ is the accretion disk temperature of the {\tt diskbb} model in units of keV; \\
$N_{\rm diskbb}$ is the normalization of the {\tt diskbb} model;\\
All errors and limits are at 90\% confidence level.
}
\end{deluxetable*}

Given the strong spectral variability during the flare and the similar spectral properties before the flare and after the flare, the $\NuSTAR$ data in the 3--79 keV band was divided into two parts: the flare spectra and non-flare spectra. The two $\Swift$/XRT observations, ObsID 00033014004 and 00033014005, from prior to the flare and after the flare, respectively (see Figure~\ref{fig:bum}), were combined with the non-flare spectra. Then, we used the model {\tt constant} $\ast$ {\tt tbabs} $\ast$ ({\tt pegpwrlw} + {\tt diskbb}) (model 1) to fit the flare spectra plus combined non-flare spectra, and freed the power-law component in these two data sets. We find that $\Swift$/XRT and $\NuSTAR$ have residuals that are not consistent with each other in the soft X-ray region where they overlap.  The residuals are also not the same for the two $\Swift$ observations.  Note that the exposure times of the $\Swift$ observations are about 1 ks (see Table~\ref{tab:obs}), much shorter than that of $\NuSTAR$, allowing for the possibility that $\Swift$ might catch short-term spectral variations in its short snapshots.  For $\NuSTAR$, the largest residuals are in the iron K$\alpha$ emission line region (Figure~\ref{fig:mo1}b, c and d) rather than in the soft X-ray band observed by $\Swift$. Therefore, in the following, we fit the $\NuSTAR$ spectra alone.

\begin{deluxetable}{clllc}
\tablewidth{\columnwidth}
\tablecolumns{5}
\setlength{\tabcolsep}{0pt}
\tabletypesize{\scriptsize}
\tablecaption{Spectral fitting of XTE J1908+094: part II \label{tab:mo2}}
\tablehead{
\colhead{Para.} & \colhead{Model 3a} & \colhead{Model 3b} &  \colhead{Model 4} & \colhead{Model 5} 
}
\startdata
 $C$ &   $0.993\pm0.003$ &  $0.993\pm0.003$  & $0.993\pm0.003$ & $0.993\pm0.003$  \\
\noalign{\smallskip}\noalign{\smallskip}
 $N_{\rm H}$ & $4.1\pm0.3$  & $4.4\pm0.3$ & $5.1^{+0.4}_{-0.3}$ & $5.3\pm0.4$  \\
\noalign{\smallskip}\noalign{\smallskip}
 $\Gamma_1$ &   $1.84^{+0.04}_{-0.06}$ &  $2.01^{+0.04}_{-0.05}$ &  $2.02\pm0.04$ & $2.02^{+0.05}_{-0.07}$ \\
\noalign{\smallskip}\noalign{\smallskip}
 $\Gamma_2$ & $2.02^{+0.04}_{-0.05}$ & $2.15^{+0.04}_{-0.05}$ & $2.16^{+0.05}_{-0.04}$ & $2.20\pm0.04$  \\
\noalign{\smallskip}\noalign{\smallskip}
 $E_{\rm fold1}$ &  $100^{\star}$ & $500^{\star}$ & $500^{\star}$ & $500^{\star}$  \\
\noalign{\smallskip}\noalign{\smallskip}
 $E_{\rm fold2}$ &  $100^{\star}$ & $500^{\star}$ & $500^{\star}$ & $500^{\star}$  \\
\noalign{\smallskip}\noalign{\smallskip}
 $N_{\rm PL1}$ &  $0.027^{+0.012}_{-0.018}$  &  $0.033^{+0.017}_{-0.024}$ & $0.036^{+0.019}_{-0.027}$ & \nodata  \\
\noalign{\smallskip}\noalign{\smallskip}
 $N_{\rm PL2}$ &  $0.05^{+0.02}_{-0.03}$ &  $0.03^{+0.05}_{-0.03}$ &  $0.06\pm0.06$ & \nodata  \\
\noalign{\smallskip}\noalign{\smallskip}
 $f_{\rm scat1}$ &  \nodata &  \nodata &  \nodata &  $0.010\pm0.010$ \\
\noalign{\smallskip}\noalign{\smallskip}
 $f_{\rm scat2}$ & \nodata & \nodata & \nodata & $0.016^{+0.010}_{-0.012}$ \\
\noalign{\smallskip}\noalign{\smallskip}
 $kT$ &  $0.719\pm0.007$ &  $0.711\pm0.007$ & $0.689^{+0.008}_{-0.012}$ & \nodata  \\
\noalign{\smallskip}\noalign{\smallskip}
 $N_{\rm diskbb}$ &  $1310^{+110}_{-100}$ & $1420^{+120}_{-110}$ & $1810^{+240}_{-160}$ & \nodata  \\
\noalign{\smallskip}\noalign{\smallskip}
 $M_{\rm BH}$ & \nodata &  \nodata & \nodata & $2.8^{+15.7}_{-0.2}$ \\
\noalign{\smallskip}\noalign{\smallskip}
 $\dot{M}$ & \nodata & \nodata &  \nodata & $1.3^{+6.6}_{-0.7}$ \\
\noalign{\smallskip}\noalign{\smallskip}
 $D_{\rm BH}$ & \nodata &  \nodata  & \nodata & $10^{\star}$ \\
\noalign{\smallskip}\noalign{\smallskip}
 $N_{\rm kerrbb}$  & \nodata &  \nodata & \nodata & $1.7^{+4.9}_{-0.8}$ \\
\noalign{\smallskip}\noalign{\smallskip}
 $\rm \xi_1$ & $5300^{+1300}_{-1600}$ & $4200^{+1300}_{-1000}$ & $9000^{+7000}_{-3000}$ & $5700^{+1200}_{-1900}$  \\
\noalign{\smallskip}\noalign{\smallskip}
 $\rm \xi_2$ & $10400\pm1700$ & $10000\pm2000$ & $19500^{+500}_{-6800}$ & $11900^{+2300}_{-1500}$ \\
\noalign{\smallskip}\noalign{\smallskip}
 $N_{\rm ref1}$ &  $1.33^{+0.16}_{-0.15}$ & $2.4^{+0.8}_{-0.7}$ & $1.2^{+0.5}_{-0.6}$ & $2.4^{+0.4}_{-0.3}$ \\
\noalign{\smallskip}\noalign{\smallskip}
 $N_{\rm ref2}$ &  $1.46\pm0.14$ & $2.4^{+0.9}_{-0.7}$ & $1.2\pm0.5$ & $2.2\pm0.3$  \\
\noalign{\smallskip}\noalign{\smallskip}
{\tt Fe/solar} &  $1.5^{\star}$ & $0.9^{+0.5}_{-0.3}$ & $4.0^{+8.4}_{-1.5}$ & $1.5^{\star}$  \\
\noalign{\smallskip}\noalign{\smallskip}
 $a$ &  \nodata & \nodata & $-0.998^{+1.9}_{-0}$ & $-0.96^{+1.63}_{-0.04}$ \\
\noalign{\smallskip}\noalign{\smallskip}
 $i$ &  \nodata & \nodata & $27^{+7}_{-4}$ & $33^{+3}_{-4}$ \\
\noalign{\smallskip}\noalign{\smallskip}
 $\chi^2/{\rm dof}$ &    2256.9/2095 & 2227.7/2094 & 2209.5/2092 & 2208.5/2092 
\enddata
\tablecomments{\\
Model 3: {\tt constant} $\ast$ {\tt tbabs} $\ast$ ({\tt reflionx\_hc} + {\tt cutoffpl} + {\tt diskbb}); 
3a: fix $\rm E_{\rm fold}=100$ keV and $\rm {\tt Fe/solar}=1.5$; 
3b: fix $\rm E_{\rm fold}=500$ keV and thaw {\tt Fe/solar}. \\
Model 4: {\tt constant} $\ast$ {\tt tbabs} $\ast$ ({\tt relconv} $\ast$ {\tt reflionx\_hc} + {\tt cutoffpl} + {\tt diskbb}). \\
Model 5: {\tt constant} $\ast$ {\tt tbabs} $\ast$ ({\tt relconv} $\ast$ {\tt reflionx\_hc} + {\tt simpl} $\ast$ {\tt kerrbb}).  \\
$E_{\rm fold1}$ and $E_{\rm fold2}$ are the folding energy of exponential rolloff for the non-flare and flare spectra in units of keV; \\
$N_{\rm PL1}$ and $N_{\rm PL2}$ are the cutoff power-law normalizations at 1 keV in photons keV$^{-1}$cm$^{-2}$s$^{-1}$; \\
$f_{\rm scat1}$ and $f_{\rm scat2}$ are the scattered fractions of the {\tt simpl} model ; \\
$M_{\rm BH}$ is the black hole mass in units of the solar mass; \\
$\dot{M}$ is the disk mass accretion rate in units of $10^{18}$ g sec$^{-1}$; \\
$D_{\rm BH}$ is the distance of the black hole in units of kpc;  \\
$N_{\rm kerr}$ is the normalization of the {\tt kerrbb} model;  \\
$\rm \xi_1$ and $\rm \xi_2$ are the ionization parameters of the {\tt reflionx\_hc} model in units of $\rm erg~cm~s^{-1}$;  \\
$N_{\rm ref1}$ and $N_{\rm ref2}$ are the normalizations of reflected spectrum ({\tt reflionx\_hc}) in units of $10^{-7}$;  \\
{\tt Fe/solar} is the abundance of iron relative to solar value;  \\
$a$ is the dimensionless black hole spin;  \\
$i$ is the inclination angle of the accretion disk in units of degree;  \\
other parameters are the same as in Table~\ref{tab:mo1}. All errors and limits are at 90\% confidence level.
}
\end{deluxetable}

As shown in Figure~\ref{fig:mo1} and \ref{fig:mo3}, a strong reflection component is apparent in the residuals of this fit (model 1), leading to a large reduced $\rm \chi^2=2709.8$ for 2099 dof (see Table~\ref{tab:mo1}). Similar to some other Galactic X-ray binaries, the reflection component is composed of an iron K$\alpha$ emission line and a broad reflection excess \citep{lig88, mil07, tom14}. The emission line feature was also detected in the 2002 outburst \citep{int02,gog04}. Following \citet{int02} and \citet{gog04}, we used the Gaussian emission line model {\tt gaussian} to fit this feature and performed fits with the neutral hydrogen column density, $N_{\rm H}$, fixed to $2.5\times 10^{22}~\rm cm^{-2}$ (model 2a). We also tested fits where $N_{\rm H}$ was a free parameter (model 2b). Adding a Gaussian significantly improves the spectral fits with $\Delta {\chi^2} \gtrsim 400$ (see Table~\ref{tab:mo1} and Figure~\ref{fig:mo3}). The unabsorbed disk flux fractions, i.e., the relative disk flux contribution to the total, unabsorbed flux in the 2--20\,keV range, are larger than 80\% for both the flare and non-flare spectra, which meet the soft state criterion of \citet{mcc06} and also confirm that the $\NuSTAR$ observation was taken in the soft state. The measurement of Gaussian line centroid, $E_{\rm cent}$, is dependent on $N_{\rm H}$. Freezing $N_{\rm H}$ at $2.5\times 10^{22}~\rm cm^{-2}$, $E_{\rm cent}$ is in the iron line region (6.4 -- 7.1 keV); leaving $N_{\rm H}$ as a free parameter, $E_{\rm cent}$ is well below this energy region. Given this, we then tested fits with $N_{\rm H}$ fixed at $4.3\times 10^{22}~\rm cm^{-2}$, the average $N_{\rm H}$ when fitting the $\Swift$ spectra in soft state with a two component model consisting of power-law and disk components (Section~\ref{sec:swi}). We obtained $E_{\rm cent1} =6.2^{+0.2}_{-0.3}$ keV and $E_{\rm cent2} =5.9\pm0.3$ keV, and the line widths $\sigma_1=1.29^{+0.16}_{-0.14}$ keV and $\sigma_2=1.51^{+0.19}_{-0.18}$ keV, respectively, for the non-flare and flare spectra, with $\chi^2/{\rm dof}=2224.1/2094$.

Instead of the Gaussian emission line model, we then used the more physical model {\tt reflionx\_hc} to fit the reflection component, and replaced the simple power-law model by a power-law with an exponential cutoff {\tt cutoffpl} (model 3). The {\tt reflionx\_hc} model is an update of the model {\tt reflionx} \citep{ros99, ros05}, which calculates the reflected spectrum from an optically thick atmosphere ionized by illuminating X-rays with a cutoff power-law spectrum. The power-law photon index of {\tt reflionx\_hc} is linked to that of {\tt cutoffpl}. Compared with {\tt reflionx}, the folding energy {\tt HighECut} in {\tt reflionx\_hc} is a free parameter also linked to that of {\tt cutoffpl}. In addition, the ionization parameter, $\rm \xi$, and the abundance of iron, {\tt Fe/solar}, extend over larger ranges in {\tt reflionx\_hc}.

When left as a free parameter, the best fit value for the exponential folding energy, {\tt HighECut}, is 500\,keV, which is the upper limit of the parameter range.  As this parameter is not well-constrained, we performed fits with {\tt HighECut} fixed at 100\,keV and 500\,keV, respectively. Moreover, we also performed fits with {\tt Fe/solar} fixed at the initial value of 1.5 and as a free parameter. Good fits with reduced $\rm \chi^2$ less than 1.08 were obtained if a reflection component was added. Changing {\tt HighECut} from 100\,keV to 500\,keV, or unfreezing {\tt Fe/solar}, other model parameters change only slightly, as seen for model 3a ($\rm \tt{HighECut}=100 keV$, $\rm {\tt Fe/solar}=1.5$) and model 3b ($\rm \tt{HighECut}=500 keV$, free {\tt Fe/solar}) in Table~\ref{tab:mo2}. Similar to the power-law photon index, the ionization parameter in the flare stage is also larger than that in the non-flare stage.  

The iron K$\alpha$ emission line may be distorted by relativistic effects; therefore, a convolution model, {\tt relconv} \citep{dau10}, was adopted to calculate relativistic smearing (model 4). The {\tt relconv} model also allows for a broken power-law emissivity function for the incident emission. Compared with other relativistic smearing models, {\tt relconv} extends the black hole spin parameter range to negative values, corresponding to a disk rotating counter to the black hole's spin.

The fits also favored a high folding energy and were performed with {\tt HighECut} fixed to 100 keV and 500 keV.  We included fits with the iron abundance free and also fixed to a value of 1.5 solar. Similarly to before, freezing {\tt Fe/solar} or changing {\tt HighECut} causes little difference in the residuals and other model parameters. The inner disk radius was set to be at the innermost stable circular orbit (ISCO), and the outer disk radius was set to 400 $r_{\rm g}$, where $r_g=GM/c^2$ is the gravitational radius. The emissivity indices were fixed at the default values, and we noted that thawing these parameters or fixing the inner emissivity index at $3<q_{\rm in}<10$ and the outer emissivity index at $0<q_{\rm out}<3$ (e.g., $q_{\rm in}=5$ and $q_{\rm out}=2$, or $q_{\rm in}=8$ and $q_{\rm out}=1$) did not improve the fits significantly (the decrease in $\rm \Delta{\chi^2}$ was less than 2.7). The best fit model is shown in Table~\ref{tab:mo2} and Figure~\ref{fig:mo3}. Adding a relativistic blurring model led to only a marginally significant improvement in $\rm \chi^2$. For the spin of the black hole, a wide range is allowed, with the full parameter range (from $-0.998$ to 0.998) being covered when all the models we used are considered. This will be discussed in Section~\ref{sec:dis}.

In order to constrain the spin of black hole, the {\tt diskbb} model was replaced by a more physical disk blackbody model, {\tt kerrbb} \citep{li05}. The model calculates the disk continuum around a Kerr black hole and fully takes the relativistic effects into account. Moreover, following previous papers \citep[e.g.,][]{tom14}, an empirical Comptonization convolution model, {\tt simpl} \citep{ste09}, which assumes that a fraction of seed photons are scattered into a power-law component, was used instead of the power-law model (model 5).

\begin{figure}
\centering
\includegraphics[width=0.48\textwidth]{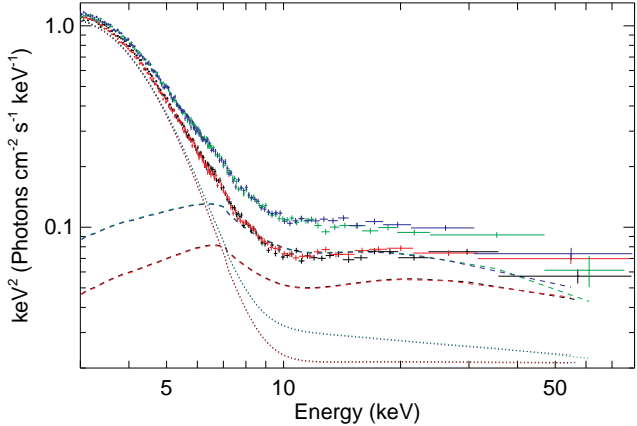} 
\caption{The unfolded $\NuSTAR$ spectra and components of model 5 in $E^2 \ast f(E)$ representation. Black and red crosses are, respectively, FPMA and FPMB spectra from the non-flare times; green and blue crosses are the flare spectra. The two upper dotted lines exhibit the {\tt simpl} $\ast$ {\tt kerrbb} components of the flare and non-flare spectra, while the two lower dashed lines exhibit the reflection components. The reflection component of the flare spectra is stronger than that of the non-flare spectra.
\label{fig:mo4}}
\end{figure}

Similar to the fits above, a high folding energy was preferred by model 5. Although we also tested the fits with $E_{\rm fold}$ fixed at 100 keV and the iron abundance left as a free parameter, we only show the spectral fitting with $E_{\rm fold}=500$ keV and {\tt Fe/solar}=1.5 in Table~\ref{tab:mo2} and Figure~\ref{fig:mo3} because there is only a slight change in the goodness of fit for other values of those parameters. The distance of XTE J1908+094 is thought to be $\sim 2-10$ kpc; thus, $D_{\rm BH}$ was set to be 2\,kpc or 10\,kpc. The spin and the inclination of {\tt kerrbb} are linked to those of {\tt relconv}. Other model parameters were fixed at the default values. We obtained a very small improvement in the fits with the reduced $\rm \chi^2$ of 1.06 for 2092 dof. Except for the normalization of the {\tt kerrbb} model, all model parameters show little changes if $D_{\rm BH}$ was changed from 10\,kpc to 2\,kpc. Thus, we only show the spectral fitting with $D_{\rm BH}=10$ kpc (see Figure~\ref{fig:mo4}). The BH spin can take values in a wide range, from $-0.998$ to $\sim 0.7$. The unabsorbed flux in the 2--12 keV band are $2.7\times10^{-9}$ \ergcms~and $2.9\times10^{-9}$ \ergcms~for the non-flare and flare spectra, respectively. Using the average flux over the non-flare and flare stages, and assuming a typical $M_{\rm BH}$ of $10~M_\sun$ and $D_{\rm BH}=(2-10)$ kpc, the source luminosity is $(1-34)\times10^{36}$ \ergs~and the Eddington fraction ($L/L_{\rm Edd}$) is 0.1\% -- 2.7\%. While the upper part of the $L/L_{\rm Edd}$ range would not be unusual for a soft state, the lower part of the range is low for a soft state \citep[e.g.,][]{yu09}, and this may favor a source distance closer to 10 kpc than 2 kpc.

Previously, using the $\BeppoSAX$ MECS spectra from the 2002 outburst, \citet{mil09} measured the spin of the black hole in XTE J1908+094. The thermal emission was not detected in these spectra; thus, they used the reflection component to constrain the spin and reported a value of $0.75\pm 0.09$. If we fix the spin at 0.75 and set the other parameters to be the same as for model 5, the quality of the spectral fit is still good. Other than the BH mass being larger, the parameters are similar to those of model 5. However, given the large uncertainties in the spin, distance and inclination, it is impossible to constrain $M_{\rm BH}$ with our current data. The inclination measurement is independent of the BH spin that we assume with a value of $\sim 30^{\circ}-40^{\circ}$, similar to $i=45^{\circ}\pm 8^{\circ}$ reported by \citet{mil09}. 

Although these models containing the disk, the power-law (Comptonization) and the reflection components fit the $\NuSTAR$ spectra well, upon closer inspection, we find a small bump in the residuals near 8--9\,keV (see Figure~\ref{fig:mo3}). A similar feature is also observed in some other $\NuSTAR$ spectra, such as Cyg X-1 \citep{tom14}. Adding a Gaussian emission line with $E_{\rm cent} \sim 8.2$ keV and $\sigma \sim 0.3$ keV, the spectral fits are improved with $\Delta {\chi^2} \sim 16$, and the key parameters change only slightly. The line feature is likely related to a combination of iron K$\beta$ and nickel emission, neither of which are included in the {\tt reflionx\_hc} model \citep{wal15}.

\section{Discussion}
\label{sec:dis}

We have presented $\NuSTAR$ and $\Swift$ observations of XTE J1908+094 during its 2013 outburst. Similar to the two previous outbursts, the source was first detected in the hard state, then went into the soft state and decayed rapidly afterwards. The time interval between the 2013 outburst and the last one is about 10 years, which is much longer than that between the two previous outbursts.

The $\NuSTAR$ light curves show a flare of $\sim$ 40 ks duration. Fitting the spectra with the two components combining model of {\tt diskbb} plus {\tt pegpwrlw} prior to, during and after the flare, we found the power-law component, rather than the disk component, exhibited major changes during the flare. The power-law was softer and brighter during the flare but seems to be stable in the stages prior to and after the flare. A possible scenario is that there was an injection of high-energy particles (perhaps due to a jet ejection or shocks in the accretion disk) during the flare; thus, the flux of the power-law component increased, and the power-law index varied. If we keep the power-law model constant during the whole observation and add another power-law model in fitting the flare spectrum, the extra emission is found to have a photon index of $\rm \Gamma =2.61\pm0.04$ with a 3--79 keV flux of $\sim 1.8\times10^{-10}$ \ergcms. 

Jet ejections are not unusual in Galactic X-ray binaries.  Other sources, such as GRS 1915+105 \citep[e.g.][]{mir94,fuc03}, GRO J1655-40 \citep[e.g.][]{tin95} and Cygnus X-1 \citep[e.g.][]{sti01}, also show ejection events. There are at least two types of ejections: one is the discrete outflow usually appearing in the hard-to-soft state transition \citep[e.g.,][]{fen04,cor04}, and the other is the compact jet occurring in the hard state and disappearing in the hard-to-soft state transition \citep[e.g.,][]{fen99,cor00, cor03}. The radio flux of XTE J1908+094 showed a significant increase between 2013 November 5 and November 6 \citep{rus13b}, which was 2--3 days before the $\NuSTAR$ observations. \citet{rus13b} and \citet{cor13} suggested that the source ejected some optically thin radio-emitting plasma during the period. In fact, the radio flux peaked during the $\NuSTAR$ observation. Also, the radio polarization measurements and the radio spectrum are consistent with the discrete ejection interpretation \citep{cur15}. Moreover, we note that radio flares are accompanied by X-ray flares in some X-ray binaries \citep[e.g.][]{wil07}. Based on the facts that the source was in a transition from the hard state to the soft state and the radio flare was apparent, such X-ray ejections would not be surprising. The X-ray flare observed by $\NuSTAR$ may have been caused by a discrete ejection.

Although a couple of faint X-ray jets were detected from microquasars a few years after the ejection \citep[e.g.,][]{cor02,cor05}, the X-ray emission produced by the plasmoid ejection may not be enough to explain the observed flux of XTE J1908+094. Thus, sudden changes of the temperature or the size of the corona may be another scenario. We used a Comptonization model {\tt comptt} \citep{tit94} to replace the power-law model in model 1, and untied different combinations of parameters in the non-flare and flare spectra. If the plasma temperature, $kT_{\rm e}$, and the 3--79 keV flux in units of $\rm 10^{-12}$ \ergcms, $N_{\rm comp}$, are allowed to be different for the different spectra, we obtain $kT_{\rm e}=500^{+0}_{-140}$\,keV and $N_{\rm comp}=377\pm4$ from the non-flare spectra, and $kT_{\rm e}=306^{+11}_{-77}$\,keV and $N_{\rm comp}=526\pm5$ during the flare, with $\chi^2/{\rm dof}=2660.5/2098$; if, instead of the temperature, the plasma optical depth parameter, $\tau$, is allowed to be free, we obtain $\tau=0.030^{+0.028}_{-0.002}$ and $N_{\rm comp}=376\pm4$ from the non-flare spectra, and $\tau=0.010^{+0.012}_{-0}$ and $N_{\rm comp}=527\pm5$ during the flare, with $\chi^2/{\rm dof}=2658.7/2098$. We note that the ejection might remove material and cause a drop in the optical depth.  If the corona is actually part of the jet, such as its base \citep{mar05}, the two explanations that we discuss (an ejection or a change in the coronal properties) might be related.

During the hard state of the 2002 outburst, a broad line feature with an average energy of $E=5.73\pm0.09$\,keV and a line width of $\sigma=1.11\pm0.31$ keV was observed by \citet{gog04}. The feature disappeared in the soft state, whereas it reappeared when the source later entered into the hard state. The flux of the line component is strongly linked to that of the power-law component; thus, \citet{gog04} suggested that the line feature might be the Fe K$\alpha$ line from the reprocessing of the hard X-ray photons by cooler material close to the central object. In order to search for the line feature over the whole 2013 outburst, we used the same model as \citet{gog04} to fit the $\Swift$ spectra. However, for most observations, the line feature is not remarkable, and the two component model containing the {\tt diskbb} and power-law components could also fit the spectra successfully (Figure~\ref{fig:para}b). This may be due to the lower throughput of $\Swift$/XRT above 6 keV, making the line feature undetectable. Moreover, in several observations of the soft state, the spectra show a possible Fe K$\alpha$ line feature, and this is further confirmed by the $\NuSTAR$ observation (see Figure~\ref{fig:mo1} and Figure~\ref{fig:mo3} where there is an iron line and hard X-ray bump). The Gaussian emission line models could fit the line feature, with model 2b (free $N_{\rm H}$) providing a better fit (see Figure~\ref{fig:mo3} and Table~\ref{tab:mo1}). If a moderate $N_{\rm H}$ of $2.5\times 10^{22}~\rm cm^{-2}$ or $4.3\times 10^{22}~\rm cm^{-2}$ is used, $E_{\rm cent}$ agrees with the energy range of iron emission; if $N_{\rm H}$ is allowed to be free, $E_{\rm cent}$ is well below this energy range, in which case the emission line may be redshifted due to the gravitational effect. Regardless of the value of $N_{\rm H}$, the line widths are about 1-2 keV, which are similar to those reported by \citet{int02} and \citet{gog04}. 

The reflection component can also be well fitted by the {\tt reflionx\_hc} model, although the folding energy is not well constrained. As would be expected due to the stronger power-law flux during the flare, the ionization parameter, $\rm \xi$, during the flare is larger than for the non-flare spectra. 

Adding a relativistic blurring model (model 4), {\tt relconv}, provides only a small improvement on the quality of the fit to the spectrum, and the parameters of model 4 and model 5 (replacing the multi-temperature disk-blackbody by the {\tt kerrbb} model), agree with those of model 3. The reflection covering fractions, calculated from the ratio in 20--40\,keV flux between the reflection and the power-law component, are 1.1--2.2 for the non-flare spectra and 1.6--3.9 for the flare spectra based on the different models. We note that the covering fractions are larger than 1, which indicates that the X-ray emission may come from closer to the black hole and the relativistic effects are stronger so that the light is gravitationally bent \citep{min04a,min04b}. Based on the {\tt relconv} parameters, all possible values for the spin of the black hole ($-0.998$ to 0.998) are allowed when the inner radius is fixed to the ISCO, which means that the spectra may be extremely blurred, with a maximal BH spin, or somewhat less blurred with a retrograde disk or with the inner disk being ionized. Following \citet{dau14}, the reflection fraction can give a lower limit on the black hole spin when assuming a lamppost geometry (i.e., a point-like corona above the spin axis of the BH); in that case a covering fraction above 1.6 implies a spin greater than 0.6.

\section{Summary and Conclusions}
\label{sec:sum}

$\NuSTAR$ and $\Swift$ observed XTE J1908+094 during its 2013 outburst. The $\Swift$ monitoring observations show that the source reached the soft state very close to the time that $\NuSTAR$ observed the source.  A flare with a duration of $\sim40$\,ks appears in the $\NuSTAR$ light curve, peaking at $\sim$40\% above the non-flare level.  When fitting the non-flare and flare spectra with two-component models, consisting of {\tt diskbb} plus {\tt pegpwrlw} or {\tt diskbb} plus {\tt comptt}, we found that the power-law component (or the Comptonization component), rather than the disk component, went through great changes during the flare. Changes of the corona, including variations of its temperature or its size, or the ejection of hot plasma, are two possible and potentially related scenarios for the flare.  A broad iron line feature with $\sigma=1-2$\,keV is observed in the $\NuSTAR$ spectrum, which motivates a spectral model that combines a thermal disk, a power-law and a reflection component, providing a good fit to the spectrum.  Although the broad iron line provides evidence for relativistic blurring of the reflection component, we are not able to constrain the BH spin in the spectral fits, and all possible spin values, from $-0.998$ to $0.998$, are allowed.  The strong reflection component requires a covering fraction in excess of 1.0, which may be explained if light bending by the BHs gravitational field enhances the flux incident on the inner disk and suggests a spin larger than 0.6.

\acknowledgments 

This work was supported under NASA contract No. NNG08FD60C and made use of data from the $NuSTAR$ mission, a project led by the California Institute of Technology, managed by the Jet Propulsion Laboratory, and funded by the National Aeronautics and Space Administration. We thank the $NuSTAR$ Operations, Software, and Calibration teams for support with the execution and analysis of these observations. This research has made use of the $NuSTAR$ Data Analysis Software (NuSTARDAS), jointly developed by the ASI Science Data Center (ASDC, Italy) and the California Institute of Technology (USA).

{\it Facility:} \facility{$\NuSTAR$}, \facility{$\Swift$}

\end{document}